\pdfoutput=1
\documentclass[sigconf,authorversion]{acmart}
\usepackage[T2A,T1]{fontenc}
\usepackage{amsmath}
\usepackage{cleveref}
\usepackage{graphicx}
\usepackage{booktabs}
\usepackage{algorithm}
\usepackage{algpseudocode}
\usepackage{caption}
\usepackage{subcaption}
\usepackage{natbib}
\usepackage{svg}
\usepackage{colortbl}
\usepackage{fancyhdr}

\author{Nicholas Boucher}
\orcid{0000-0002-5674-3730}
\affiliation{%
  \institution{University of Cambridge}
  \country{}
  \city{}}
\email{nicholas.boucher@cl.cam.ac.uk}

\author{Luca Pajola}
\affiliation{%
  \institution{University of Padua}
  \country{}
  \city{}}
\email{luca.pajola@phd.unipd.it}

\author{Ilia Shumailov}
\affiliation{%
  \institution{University of Oxford}
  \country{}
  \city{}}
\email{ilia.shumailov@chch.ox.ac.uk}

\author{Ross Anderson}
\orcid{0000-0001-8697-5682}
\affiliation{%
 \institution{University of Cambridge \& Edinburgh}
  \country{}
  \city{}}
\email {ross.anderson@cl.cam.ac.uk}

\author{Mauro Conti}
\affiliation{%
  \institution{University of Padua}
  \country{}
  \city{}}
\email{conti@math.unipd.it}

\title[Boosting Big Brother: Attacking Search Engines with Encodings]{Boosting Big Brother:\texorpdfstring{\\}{ }Attacking Search Engines with Encodings}
\date{}

\hypersetup{
    pdftitle={Boosting Big Brother: Attacking Search Engines with Encodings},
    pdfauthor={Nicholas Boucher, Luca Pajola, Ilia Shumailov, Ross Anderson, and Mauro Conti}
}



\copyrightyear{2023}
\acmYear{2023}
\setcopyright{rightsretained}
\acmConference[RAID '23]{The 26th International Symposium on Research in Attacks, Intrusions and Defenses}{October 16--18, 2023}{Hong Kong, Hong Kong}
\acmBooktitle{The 26th International Symposium on Research in Attacks, Intrusions and Defenses (RAID '23), October 16--18, 2023, Hong Kong, Hong Kong}\acmDOI{10.1145/3607199.3607220}
\acmISBN{979-8-4007-0765-0/23/10}

\definecolor{Green}{rgb}{0.195, 0.656, 0.320}
\creflabelformat{equation}{#2#1#3}

\newcommand{\cyrillicH}{{\fontencoding{T2A}\selectfont Н}}
\newcommand{\eg}{\textit{e.g\@.}}

\newcommand{\ie}{\textit{i.e\@.}}

\fancypagestyle{firststyle}
{
   \fancyhf{}
   \chead{\raisebox{0.25in}[0in][0in]{\textit{\rule[0.5ex]{8em}{0.5pt} In 26th Symposium on Research in Attacks, Intrusions and Defenses \rule[0.5ex]{8em}{0.5pt}}}}
}

\begin{document}

\begin{abstract}
Search engines are vulnerable to attacks against indexing and searching via text encoding manipulation. By imperceptibly perturbing text using uncommon encoded representations, adversaries can control results across search engines for specific search queries. We demonstrate that this attack is successful against two major commercial search engines - Google and Bing - and one open source search engine - Elasticsearch. We further demonstrate that this attack is successful against LLM chat search including Bing's GPT-4 chatbot and Google's Bard chatbot. We also present a variant of the attack targeting text summarization and plagiarism detection models, two ML tasks closely tied to search. We provide a set of defenses against these techniques and warn that adversaries can leverage these attacks to launch disinformation campaigns against unsuspecting users, motivating the need for search engine maintainers to patch deployed systems.
\end{abstract}

\maketitle
\thispagestyle{firststyle}

\section{Introduction}\label{sec.intro}

Disinformation is a recurring threat to society exacerbated by the internet. While global internet connectivity democratizes knowledge by giving broad access to information, it also builds an easily scalable platform that can be used to provide purposefully manipulated content for promoting an adversarial goal. When adversaries coordinate the promotion of knowingly false information as truth via online platforms, we refer to such efforts as disinformation campaigns.

Search engines play a critical role in mitigating disinformation campaigns. Ethically, it would be dubious for search engines to identify and suppress disinformation; this quickly begins to take the form of censorship. However, users do expect search engines to produce representative results. From this assumption, it should be difficult for an adversary to severely manipulate search results without controlling the majority of relevant online content.

In this paper, we will show that this assumption is false. Adversaries can leverage atypical text encodings to manipulate search results in a targeted fashion.

Consider the dystopian world of George Orwell's \textit{1984}: in the novel, the enemy and ally states in an ongoing war switch roles partway through the narrative. When this occurs, the relevant government rewrites wartime propaganda to state that the previous ally had always been the enemy~\cite{orwell1984}. In such a world, we would expect that search engines would surface at least some of the plethora of historical documents representing factual history rather than surfacing an exclusive subset of documents aligned with the advertised political narrative. Using the techniques presented in this paper, though, that expectation may not hold.

Search engines, like most text processing systems, understand text according to its binary encoding. For every instance of encoded text, there exist alternative representations that render the same glyphs encoded differently. Search engines that fail to correlate different representations of the same rendered text are subject to adversarial manipulation against both indexing and searching.

Using these techniques, an adversary can provide search terms that return targeted results across search engines and prevent content from being indexed as expected. Adversaries can leverage these attacks to boost disinformation campaigns by deceiving users into believing false claims are broadly supported by search results. These techniques could also be used to adversarially limit the discoverability of legal documents such as court disclosures or patents.

\begin{figure}
     \centering
     \begin{subfigure}[b]{0.45\linewidth}
         \centering
         \includegraphics[width=\textwidth]{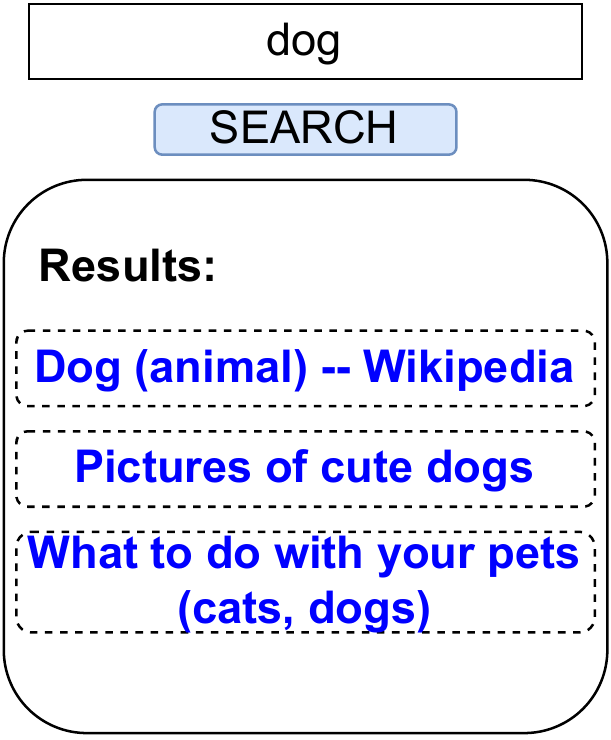}
         \caption{Benign query.}
         \label{fig:ex-ben}
     \end{subfigure}
     \hfill
     \begin{subfigure}[b]{0.45\linewidth}
         \centering
         \includegraphics[width=\textwidth]{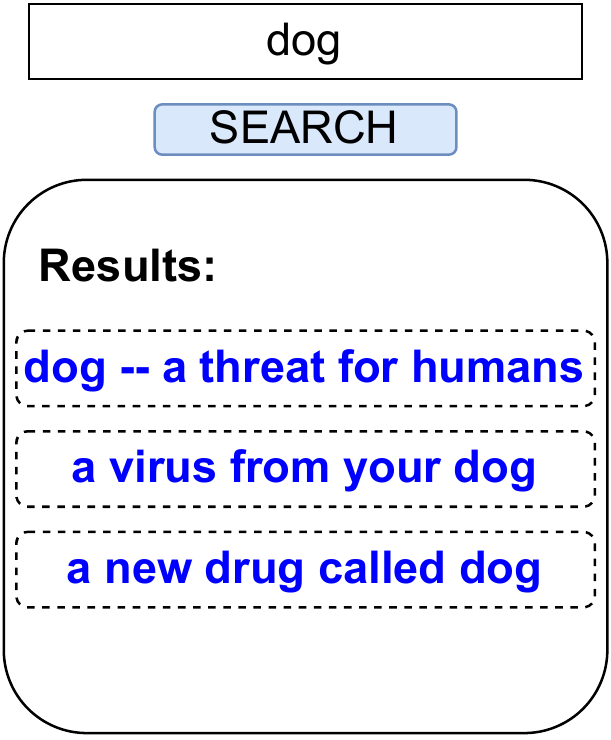}
         \caption{Malicious query.}
         \label{fig:ex-mal}
     \end{subfigure}
    \caption{An example of the attacker's goal. The figures show the search engine's top results for two ``dog'' queries, where one is benign (left) and the other adversarial (right). The queries appear identical, but the adversarial query is written with homoglyphs (\texttt{U+501} \& \texttt{U+3BF}).}
    \label{fig:example-attack}
\end{figure}

In this work, we study how malicious actors can manipulate the indexing and retrieval of web information through text encoding manipulations. We first focus on how search engines can be manipulated by analyzing the ability of attackers to perform: (1) \textit{hiding}, \ie~the ability to hide adversarial content from benign query results; and (2) \textit{surfacing}, \ie~the ability to yield adversarial results for perturbed queries. Using our techniques, an attacker may be able to publish content indexed by search engines that only appears in the search results of imperceptibly perturbed, adversarial queries. An example is shown in \Cref{fig:example-attack}, where malicious content appears only when users search using a poisoned query. Our experiments analyze how distinct search engines -- commercial (e.g., Bing and Google) and open source (Elasticsearch) -- behave under this novel threat. Furthermore, we assess how different machine learning systems that commonly support search engines can be affected by our proposed attack. In particular, we analyze the effect of our attacks on Bing's integration with GPT-4, Google's Bard model, text summarization models, and plagiarism detection models. 

Our contributions can be summarized as follows:
\begin{itemize}
    \item We introduce a novel attack against search engine indexing and querying through adversarial text encodings.
    \item We conduct experiments demonstrating that these attacks are successful against three major commercial and open source search engines.
    \item We demonstrate that these attacks also extend to chatbot search assistants, text summarization models, and plagiarism detection models, adding further attacks relevant to modern search engine ecosystems.
    %\item We propose defenses that search engine maintainers can use to mitigate these attacks.
\end{itemize}
\section{Background}\label{sec.bad-characters}

\subsection{Encoding Text}

Within computing systems, text must be represented in some binary fashion. To accomplish this, software designers first select a text specification such as Unicode~\cite{unicode_2022}\footnote{Unicode is perhaps the most common modern text specification. According to \href{https://w3techs.com/technologies/details/en-utf8}{w3techs.com/technologies/details/en-utf8}, 97.9\% of observed websites use the UTF-8 Unicode encoding.}. This specification maps characters to numbers known as code points\footnote{When written, Unicode code points are typically formatted as preceding with \texttt{U+}.}. These code points are then represented in binary according to a chosen encoding, such as UTF-8.

When text is rendered, binary values interpreted as code points according to the chosen encoding code points are resolved to characters via the text specification, then any control characters implementing special behavior are resolved, and finally characters are rendered as glyphs using the chosen font.

\subsection{Imperceptible Perturbations}

Rendered text is not injective with encoded representations; that is, identical rendered text can be represented by many different encoded values within the same text specification.

The following categories represent variations of imperceptible perturbations, also known as bad characters, which can be used to encode the same visual text with different values~\cite{boucher2022bad}:

\begin{enumerate}
    \item \textbf{Invisible Characters}\\
    Invisible characters are characters that render to the absence of a glyph. They often exist for formatting purposes in specific contexts, and have no effect on text outside of those contexts. One example in Unicode is the Zero Width Space (ZWSP). Invisible characters can be injected arbitrarily to perturb text encodings without visual effect.\\
    \item \textbf{Homoglyphs}\\
    Homoglyphs are characters that render to the same or nearly the same glyph. An example in Unicode is the Latin H and the Cyrillic \cyrillicH. Homoglyph characters can be substituted with each other to generate imperceptible perturbations.\\
    \item \textbf{Reorderings}\\
    Reorderings exploit bidirectional text support by changing the display order of text while also changing the encoded order of text to offset the effect. An example in Unicode can be implemented using the Right-to-Left Override (RLO) control character. Since encoded values are manipulated without changing the rendered text, reorderings are a form of imperceptible perturbation.\\
    \item \textbf{Deletions}\\
    Deletions exploit control characters designed to remove adjacent text by injecting arbitrary characters followed by an equivalent number of deletion control characters. An example in Unicode is the Delete (DEL) character. Deletions are powerful because they allow injecting characters into text without rendering them, although deletions are not imperceptible in all settings.
    
\end{enumerate}
\section{The Attack}

\begin{table*}[t]
\caption{Perturbation Techniques Used in Bad Search Wiki}
\label{tbl:perturbations}
\begin{tabular}{@{}lll@{}}
\toprule
\textbf{Perturbation Name} & \textbf{Category}            & \textbf{Description}                                                              \\ \midrule
base              & Unperturbed         & Text without perturbation to serve as a control.                         \\
zwsp              & Invisible Character & Injects a Zero Width Space between all adjacent characters.              \\
zwnj              & Invisible Character & Injects a Zero Width Non-Joiner between all adjacent characters.         \\
zwj               & Invisible Character & Injects a Zero Width Joiner between all adjacent characters.             \\
homo              & Homoglyph           & Substitutes each character with a randomly chosen homoglyph.             \\
rlo               & Reordering          & Wraps text with a Right-to-Left Override and reverses logical order.     \\
bksp              & Deletion            & Injects an \texttt{X} followed by a backspace character (\texttt{U+8}).  \\
del               & Deletion            & Injects an \texttt{X} followed by a backspace character (\texttt{U+7F}). \\ \bottomrule
\end{tabular}
\end{table*}

\subsection{Threat Model}

We propose a threat model in which an adversary seeks to insert web pages as highly ranked results for a specific search engine query executed by a victim user, where highly ranked is defined as appearing on the first default-sized page of results. The adversary does not have the ability to modify the search engine, nor does the adversary have knowledge of which search engine will be used by the victim. The adversary can create public websites that will be indexed by the search engine, but does not have the ability to promote those sites within the search engine index above other similar sites which they do not control.

A practical realization of this adversary is an actor conducting a disinformation campaign. This actor wishes for their search results to be prioritized over other results such that evidence conflicting the disinformation is crowded out by content under the actor's control.

\subsection{Attack Technique}

Within our threat model, an attacker can perform this attack by leveraging imperceptible perturbations.

Barring defenses, search engines understand both indexed content and search queries according to their encoded values. That is, visually identical text with and without imperceptible perturbations will be interpreted by search engines as distinct values. Adversaries can leverage this to plant poisoned content within the search engine index, and then surface that content to victim users who search using the imperceptibly perturbed string. This content will also be unlikely to show up in unperturbed queries, meaning that poisoned content is shown primarily to targeted users.

To illustrate this attack, consider the following example:
\begin{enumerate}
    \item Eve is running a disinformation campaign to deceive victims into believing that an unproven drug is an effective treatment for a certain ailment. Eve creates multiple fake websites attesting to the efficacy of the drug.
    \item Eve then modifies these sites such that each occurrence of the drug's name is imperceptibly perturbed with the same perturbation.
    \item Eve submits her sites for indexing in multiple major commercial search engines.
    \item Once Eve validates that the sites are indexed, she publicizes the drug on social media platforms using the perturbed version of the name.
    \item Alice, a victim, sees Eve's social media post and searches her favorite search engine to learn more about the drug. She copies the name of the drug from the social media post into the search bar rather than retyping the long name.
    \item Without realizing, Alice has searched for the imperceptibly perturbed version of the drug's name. The search engine returns Eve's fake websites as the top results, since they are the only indexed sites containing the search term imperceptibly perturbed in that manner.
    \item Alice is now deceived into believing that most internet results support Eve's disinformation claims.
\end{enumerate}

By leveraging such techniques at scale, an adversary can significantly promote search engine results to support a broader disinformation campaign.

We note that in this setting, it is not necessary to deceive all potential victims. Some users may retype search queries thereby removing imperceptible perturbations, and yet others may go directly to trusted sources of information rather than search engines. So long as a subset of users leverage copy+paste or click-to-search functionality and review only "top" ranked sources, this attack will have victims.
\section{Evaluation}

\subsection{Methodology}
Our experiments test whether search engines can be affected by the presence of imperceptible perturbations both in indexing, \ie~parsing crawled content, and querying, \ie~performing searches. 
We define three distinct measures for evaluating the impact of imperceptible perturbations on search engines:
\begin{itemize}
    \item \textit{Disruption Potential}, measuring the SERP mismatch between benign and perturbed queries;
    \item \textit{Hiding Potential}, measuring the ability of perturbed pages being discovered through benign queries;
    \item \textit{Surfacing Potential}, measuring the ability of perturbed pages being discovered through perturbed queries. 
\end{itemize}
Disruption is a broad measurement of whether a search engine is affected by imperceptible perturbations. Hiding is a more specific metric that determines whether indexed content can be withheld from search results for typical users of a search engine, while surfacing determines whether targeted content can be surfaced in search results for a targeted users. An fully vulnerable platforms has all of these properties. In the following paragraphs, we define these measures formally.

\paragraph{Disruption Potential}
We analyze the discrepancy in Search Engine Results Pages (SERPs) between benign queries and their imperceptibly perturbed counterparts. In particular, SERP $S$ from engine $e$ is nothing more than a list of URLs $u$ representing the highest ranked results for the given query $x_i$:
\begin{equation}\label{eq:serp}
    S_e(x_i) = [u_0, u_1, ..., u_n],
\end{equation}
Therefore, we compare the SERP of $x_i$ and $x_i^{adv}$ as follows:
\begin{equation}\label{eq:similarity}
    M_d(S_e, x_i, x_i^{adv}) = 1- \frac{|S_e(x_i) \cap S_e(x_i^{adv})|}{|S_e(x_i)|},
\end{equation}
where $|\cdot|$ is the cardinality of the set. 
In other words, we attempt to measure how many correct results $S_e(x_i^{adv})$ contains. $M_d$ is defined in $[0, 1]$, where 1 indicates that $S_e(x_i) \cap S_e(x_i^{adv}) = \emptyset$, \ie~there is a total mismatch between $S_e(x_i)$ and $S_e(x_i^{adv})$; conversely, when $M_d = 0$, the search engine $S_e$ is not affected by the perturbation, \ie~$S_e(x_i)$ = $S_e(x_i^{adv})$. 

\paragraph{Hiding Potential}
We analyze the ability of an attacker to hide content from a search engine's index using the hiding score. For this metric, we define $u_i^{adv}$ as a URL containing imperceptibly perturbed content relevant to the unperturbed query $x_i$. We, therefore, define the hiding metric as follows:
\begin{equation}
M_h(S_e, x_i, u_i^{adv}) =
\begin{cases} 
0 & \text{if} \; u_i^{adv} \in S_e(x_i), \\ 
1 & \text{otherwise}.  
\end{cases}
\end{equation}
Intuitively, this means that a high $M_h$ score implies an attacker can prevent content from appearing in typical search results by adding imperceptible perturbations.

\paragraph{Surfacing Potential}
Similarly, we analyze the ability of an attacker to surface a specific page in search engine results for a query of their choice using the surfacing score. For this metric, we define $u_i^{adv}$ as a URL containing imperceptibly perturbed content relevant to the perturbed query $x_i^{adv}$. We therefore define the surfacing metric as follows:
\begin{equation}
M_s(S_e, x_i^{adv}, u_i^{adv}) =
\begin{cases} 
1 & \text{if} \; u_i^{adv} \in S_e(x_i^{adv}), \\ 
0 & \text{otherwise}.  
\end{cases}    
\end{equation}
Intuitively, this means that a high $M_s$ score implies an attacker can surface imperceptibly perturbed content with high confidence for a given perturbed query.

\begin{figure}[t]
    \centering
    \includegraphics[width=\columnwidth]{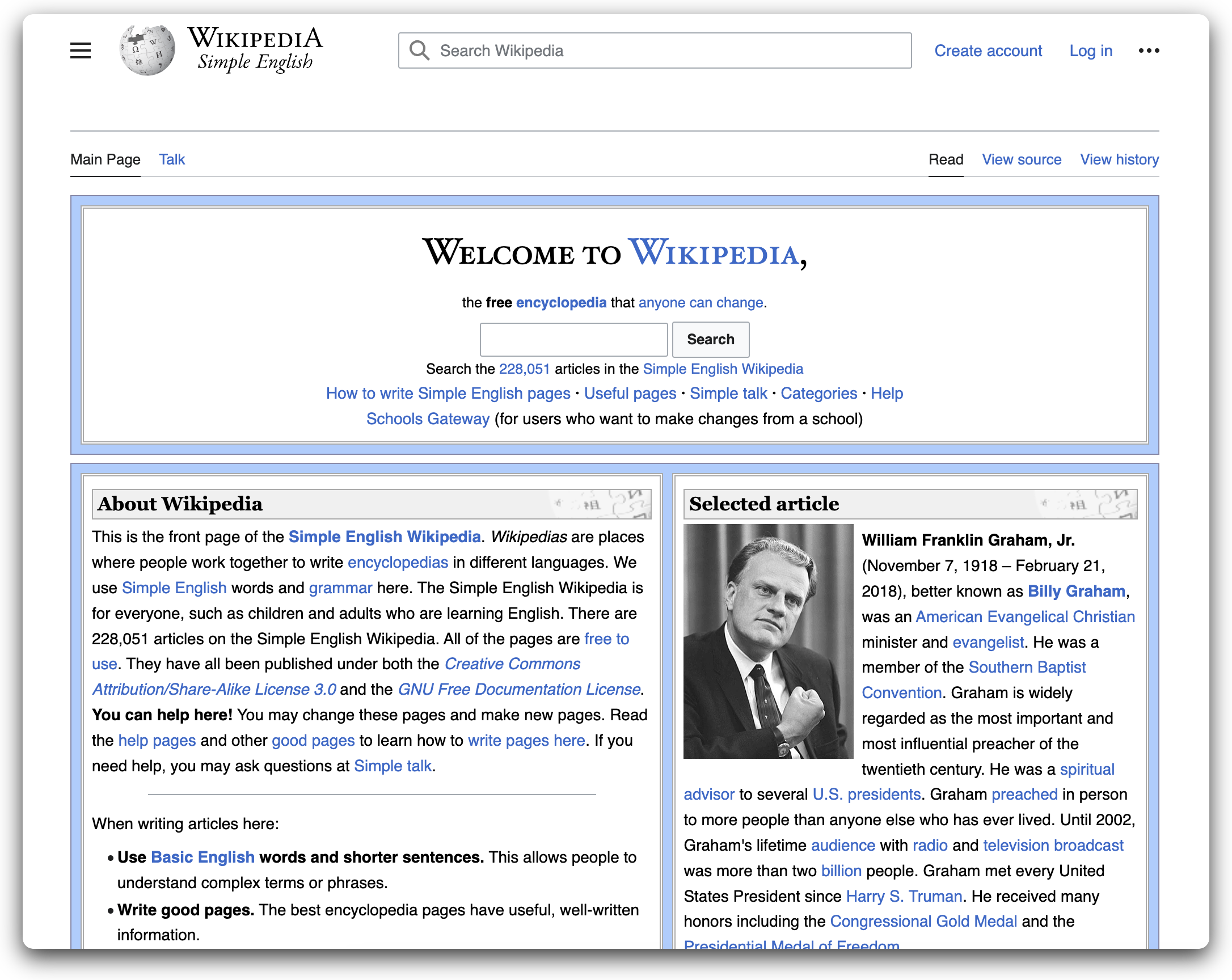}
    \caption{Simple Wikipedia.}
    \label{fig:simple-wikipedia}
\end{figure}

\subsection{Experimental Setup}

We evaluate our attack on three common search engines: Google, Bing, and Elasticsearch. Of these, Google and Bing, the two most common commercial search engines~\cite{statcounter_2022}, are both black-box systems, while Elasticsearch is an open source search engine implementation.

\begin{figure}[t]
    \centering
    \includegraphics[width=\columnwidth]{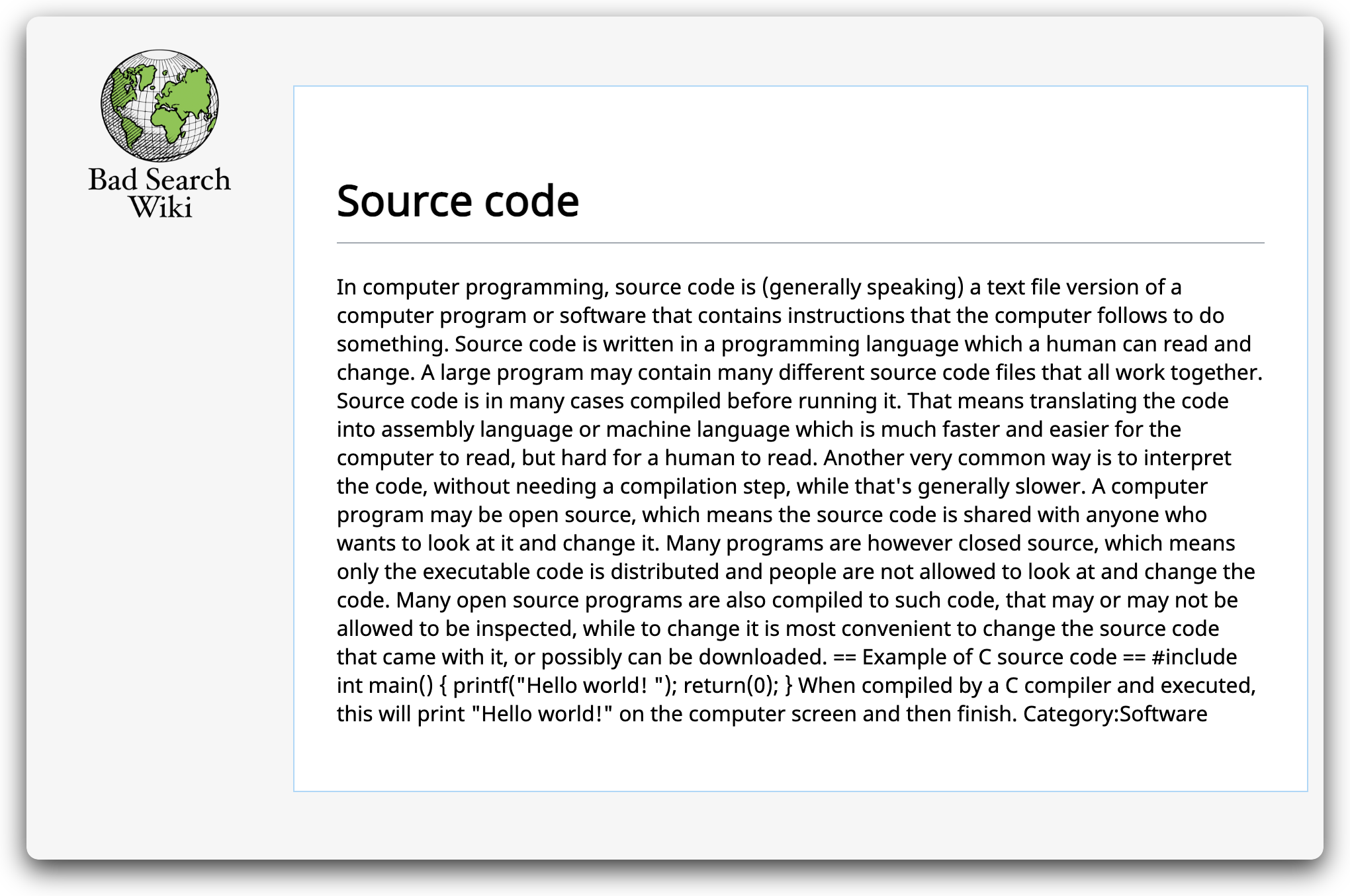}
    \caption{Bad Search Wiki.}
    \label{fig:bad-search-wiki}
\end{figure}

Our evaluation analyzes search results on an imperceptibly perturbed version of Simple Wikipedia\footnote{\href{https://simple.wikipedia.org}{simple.wikipedia.org}}. Pictured in \Cref{fig:simple-wikipedia}, Simple Wikipedia is an English-language instance of Wikipedia aimed at children and adults learning the language. At the time of writing, it contained 224,219 articles, making it more conducive for experimentation than the significantly larger primary Wikipedia instance. We leveraged eight perturbations for our experiments representing all four categories of imperceptible perturbations: invisible characters, homoglyphs, reorderings, and deletions. These perturbations are described in \Cref{tbl:perturbations}. We note that from our testing the deletion techniques produce visual artifacts in most web browsers; we include these techniques for robustness, but they would likely be avoided in practice. We also note that \textit{base} is the name given to the control setting in all of our experiments for which no perturbations are applied.

Experiments against Elasticsearch involved running a local instance of the search engine and indexing the entirety of Simple Wikipedia for each perturbation technique.

Experiments against Google and Bing were more complicated, as we could not programmatically specify free-form data for indexing. To get around this, we deployed a mirror of Simple Wikipedia with added perturbations to a web server under our control\footnote{\href{https://badsearch.soc.srcf.net}{badsearch.soc.srcf.net}}. We then requested indexing of the site with both Google and Bing, and leveraged each search engine's API for querying the index of our site only. It was rather challenging to get these sites into the index, requiring properly formatted sitemaps, robot files, crawl requests, and a friendly domain name before a sufficient portion of the site was indexed by either search engine. To accommodate indexing constraints, we randomly selected 100 articles for perturbations across our eight techniques for experiments with Google and Bing rather than using the entirety of Simple Wikipedia.

Our experimental online wiki -- dubbed the ``Bad Search Wiki'' -- for which we depict a sample article in \Cref{fig:bad-search-wiki} displays the title and article text taken from an export of Simple Wikipedia. We remove all formatting and embedded media from each article. Each article is repeated 8 times within the site, with each instance having a different perturbation applied. URLs do not contain any article-specific information not already in the page to prevent any effect on the indexing process.

Following our evaluations of Google, Bing, and Elasticsearch, we provide one additional set of experiments for Google and Bing that query the open internet rather than the Bad Search Wiki alone. This set of experiments aims to complement the Bad Search Wiki evaluations to show that the results presented throughout this paper also apply to web properties outside of our control.

The source code for our Bad Search Wiki and each experiment is available on GitHub\footnote{\href{https://github.com/nickboucher/search-engine-attacks}{github.com/nickboucher/search-engine-attacks}}.

\begin{figure}[t]
    \centering
    \includesvg[width=\columnwidth]{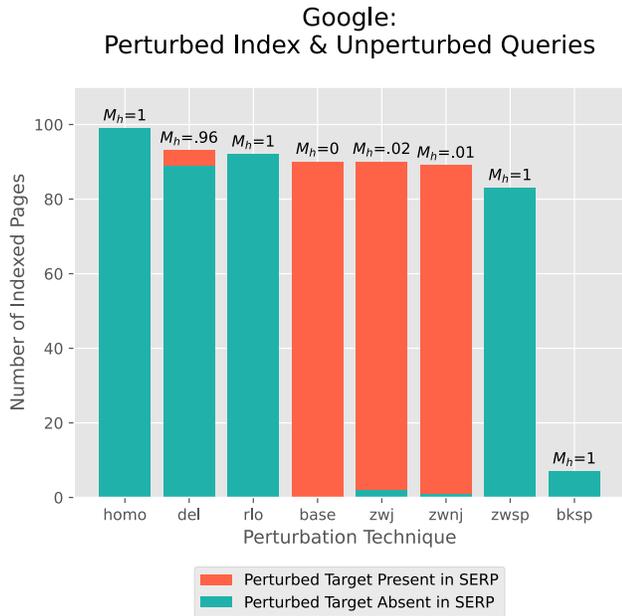}
    \caption{Google Hiding Experiment. The higher $M_h$, the stronger the attack. Green represents attack success.}
    \label{fig:google-hiding}
\end{figure}

\subsection{Google}

Google offers a Programmable Search Engine product\footnote{\href{https://developers.google.com/custom-search}{developers.google.com/custom-search}} that allows automated querying of the Google search index. The API allows specifying individual URL patterns for inclusion in each search query, allowing us to select which perturbation technique pages to query with each search. Google also offers a Search Console\footnote{\href{https://search.google.com/search-console}{search.google.com/search-console}} that makes is easy to detect if pages that you own are included in Google's index.

We conducted two different experiments against Google using the Bad Search Wiki which we will describe below.

The first experiment, which we call the \textit{hiding experiment}, tests whether Google's search engine displays different behavior between the same query in perturbed and unperturbed form. To do this, we query the subset of Google's index that contains only articles perturbed using a single perturbation technique, such as zwsp, on our experimental site. For the search query, we use the unperturbed name of the target article. If Google returns the target article in its perturbed form, we can conclude that Google removes this form of perturbation during indexing.

\begin{figure}[t]
    \centering
    \includesvg[width=.9\columnwidth]{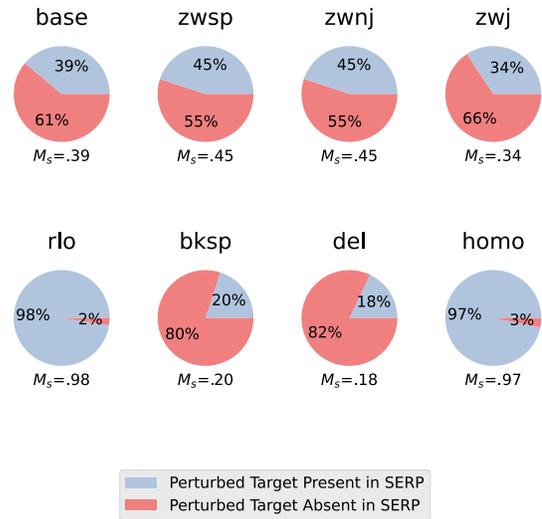}
    \caption{Google Surfacing Experiment. The higher $M_s$, the stronger the attack. Blue represents attack success.}
    \label{fig:google-surfacing}
\end{figure}

The results of the hiding experiment are shown in \Cref{fig:google-hiding}. We note that despite spending a full year attempting to get Google to index the entire Bad Search Wiki, there were some pages that were never included in the index. Missing pages are represented by shorter bars in this visualization. Likewise, pages that were indexed and returned the target article in its perturbed form in the first SERP, \ie~the top 10 URL hits, are represented in red; these represent pages that were not successfully ``hidden'' from the search engine through perturbations. Indexed target pages that were not returned in the first results page are represented in green, as these represent pages that were successfully ``hidden''. Mean $M_h$ values are reported for each technique, where $S_e(x_i)$ is defined as the singleton set including only the target page perturbed using the selected technique.

Unsurprisingly, we see that unperturbed queries against an unperturbed index always include the target result. More surprising, however, is that ZWJ and ZWNJ results are also correctly included the majority of the time. This result suggests that Google is robust against ZWJ and ZWNJ, but not against ZWSP, homoglyphs, DELs, BKSPs, and RLOs.

The second experiment, which we call the \textit{surfacing experiment}, queries the entire Bad Search Wiki site (all perturbations) with the search query being the perturbed form of an article title. If the same perturbed form of the article is present in the first page of query results, we can conclude that the perturbation technique is a good candidate for our attack when used against Google.

From the results, we can see that RLOs and homoglyphs are particularly good techniques for targeted content poisoning on Google. It is somewhat surprising that unperturbed (base) queries aren't 100\% present, but we suspect that, following the results from the hiding experiment, Google views base, ZWJ, and ZWNJ as duplicative content, and randomly selects only one of these pages to show in the top search results.

From these results, we can conclude that reordering and homoglyph perturbations are highly effective attack techniques against Google. We can also conclude that Google has existing mitigations in place preventing ZWJ and ZWNJ-based perturbation attacks.

\subsection{Bing}

We conducted the same set of experiments against Bing that we conducted against Google and represent results equivalently. Bing also provides programmatic search engine querying via its Custom Search API product\footnote{\href{https://www.customsearch.ai}{customsearch.ai}}. Similarly, web property owners can request and validate Bing indexing using Webmaster Tools\footnote{\href{https://www.bing.com/webmasters}{bing.com/webmasters}}.

\begin{figure}[t]
    \centering
    \includesvg[width=\columnwidth]{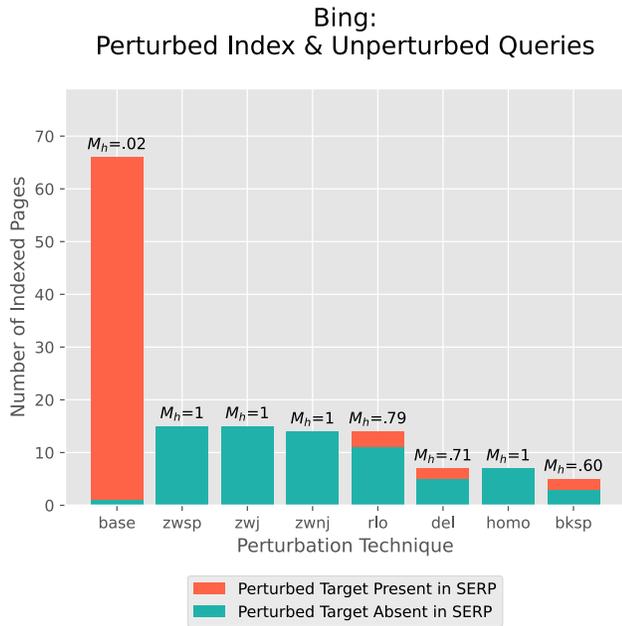}
    \caption{Bing Hiding Experiment. The higher $M_h$, the stronger the attack. Green represents attack success.}
    \label{fig:bing-hiding}
\end{figure}

The first experiment conducted against Bing was likewise the hiding experiment. As with Google, in this experiment we searched the index only including the target perturbation using the unperturbed article title as the search query. We note that compared to Google, we found it very difficult to index a large number of pages in Bing. The number of indexed pages is lower, despite the fact that we launched a second instance of the Bad Search Wiki site and combined the Bing results data for both.

The results shown in \Cref{fig:bing-hiding} suggest that Bing views each perturbation technique distinctly; results and queries are not correctly associated between perturbed/unperturbed version, with the exception of the unperturbed control which was highly discoverable as expected. This data implies that there are not likely to be defenses built into Bing for any of the measured perturbation techniques.

The second experiment conducted against Bing likewise was the surfacing experiment. As with Google, in this experiment we searched the entire Bad Search Wiki index (all perturbations) using perturbed article titles.

\begin{figure}[t]
    \centering
    \includesvg[width=.9\columnwidth]{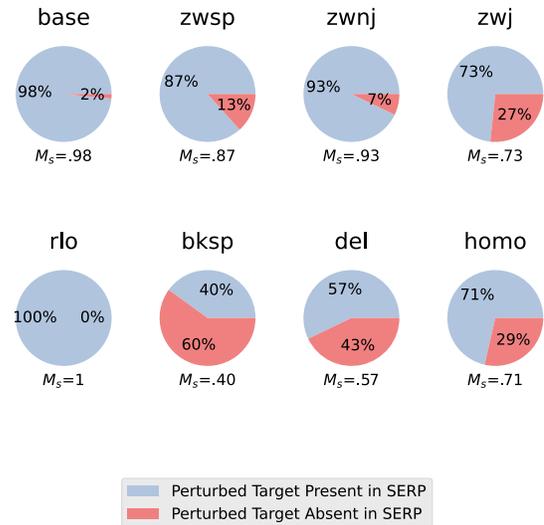}
    \caption{Bing Surfacing Experiment. The higher $M_s$, the stronger the attack. Blue represents attack success.}
    \label{fig:bing-surfacing}
\end{figure}

The results shown in \Cref{fig:bing-surfacing} further suggest that Bing has little to no mitigations in place for perturbation attacks. The data are not as binary for each technique, but we suspect that the noise is higher in this experiment due to the smaller sample size per technique from indexing limitations. From this data, though, we can conclude that RLOs, ZWNJs, and ZWSPs are highly effective attack techniques against Bing. The remaining techniques are also likely to represent effective attacks with slightly lower attack success rate (ASR).

\subsection{Elasticsearch}

Since Elasticsearch is an open-source search engine, there is no need to deploy websites and request indexing to run experiments; rather, we can simply directly index our perturbed article titles and content. Indexing ability is thus not a limiting factor, and we therefore indexed all pages in Simple Wikipedia for each perturbation technique. We also added two additional perturbation techniques not seen in our previous experiments: \texttt{zwsp2}, for which article titles are alone perturbed with a random number of ZWSPs held constant for repeated words, and \texttt{homo2} in which homoglyphs were manually selected to minimize visual perturbation artifacts rather than randomly selecting homoglyph substitutions.

Our experiments are conducted against Elasticsearch 8.5.3 run via Docker\footnote{\href{https://hub.docker.com/_/elasticsearch}{hub.docker.com/\_/elasticsearch}}. We performed queries via the Python Elasticsearch client.

Although the experimental setup was slightly different, we performed the same two evaluations as those conducted against Google and Bing and represent results equivalently.

\begin{figure}[t]
    \centering
    \includesvg[width=\columnwidth]{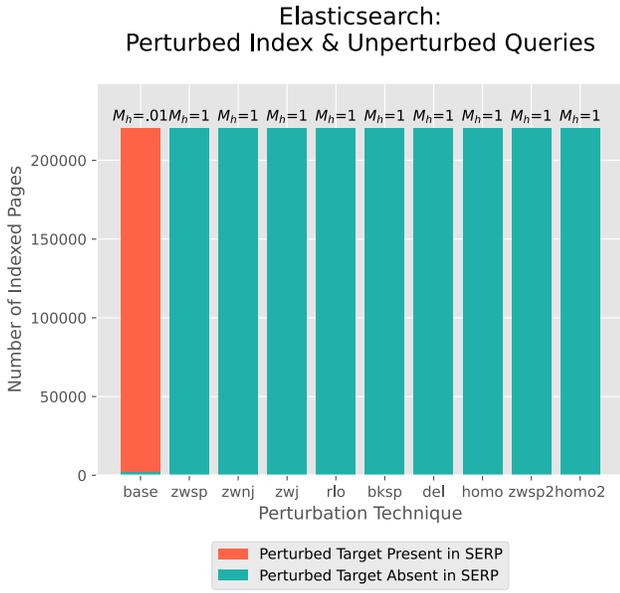}
    \caption{Elasticsearch Hiding Experiment. The higher $M_h$, the stronger the attack. Green represents attack success.}
    \label{fig:elastic-hiding}
\end{figure}

The first experiment was therefore the hiding experiment, for which the results can be found in \Cref{fig:elastic-hiding}. The results suggest that Elasticsearch interprets each perturbation technique as a distinct value from the unperturbed equivalent. 

\begin{figure}[t]
    \centering
    \includesvg[width=.9\columnwidth]{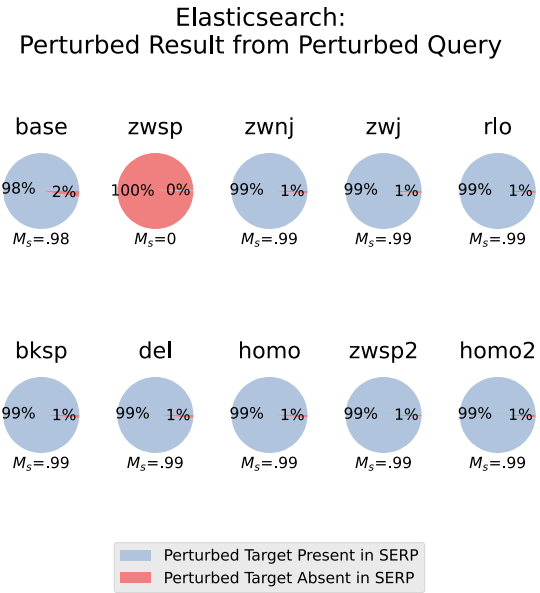}
    \caption{Elasticsearch Surfacing Experiment. The higher $M_s$, the stronger the attack. Blue represents attack success.}
    \label{fig:elastic-surfacing}
\end{figure}

The second experiment was thus the surfacing experiment, and the results can be found in \Cref{fig:elastic-surfacing}. The results indicate that all techniques other than zwsp are highly successful in surfacing targeted, perturbed content using similarly perturbed search queries. This implies that ZWSPs are ignored in search queries, but all other perturbation techniques are treated distinctly from their unperturbed counterparts.

\subsection{Open-Internet Measurement}\label{sec.CS1}

In a final set of experiments, we evaluated the performance of Google and Bing over the general internet with perturbed queries. Each engine was queried using the official API~\cite{google-api, microsoft-api} with queries formed from the question/answer dataset provided by the \textit{boolq} version of the \textit{super-glue} dataset~\cite{clark2019boolq,wang2019superglue}. From this dataset, we randomly selected 100 questions and injected imperceptible perturbations in random positions. In our experiments, we vary the number of injected characters as follows: $\{1, 3, 5, 7, 9\}$. We then measure the \textit{Disruption Potential} as defined in \Cref{eq:similarity}. This set of experiments seeks only to validate the performance of search engines against imperceptible perturbations in the most general, open-internet setting, and does not seek to determine whether the URLs returned are related to malicious campaigns.

\begin{table}[b]
    \centering
    \caption{Top 10 web pages occurring among perturbed URLs.}
    \begin{tabular}{lr} \toprule
         \textbf{Website} & \textbf{Occurrences}  \\ \midrule
         en.wikipedia.org & 1351 \\
         www.researchgate.net & 1132 \\
         www.quora.com & 449 \\
         www.imdb.com & 366 \\
         www.ncbi.nlm.nih.gov & 353 \\
         www.coursehero.com & 293 \\
         www.youtube.com & 292 \\
         screenrant.com & 198 \\
         archive.org & 196 \\
         www.nytimes.com & 196 \\\bottomrule
    \end{tabular}
    \label{tab:common-webpages}
\end{table}

% \begin{figure*}[t]
%     \centering
%     \includegraphics[width=0.75\linewidth]{figures/graphics/search-engine-queries.pdf}
%     \caption{Effect of imperceptible perturbations on different search engines at the varying amount of perturbation: the lower the similarity, the stronger the attack. The higher $M_d$, the stronger the attack. \todo{i'll fix this tomorrow morning}}
%     \label{fig:cs1-results}
% \end{figure*}

\begin{figure*}
     \centering
     \begin{subfigure}[b]{0.45\linewidth}
         \centering
         \includegraphics[width=\textwidth]{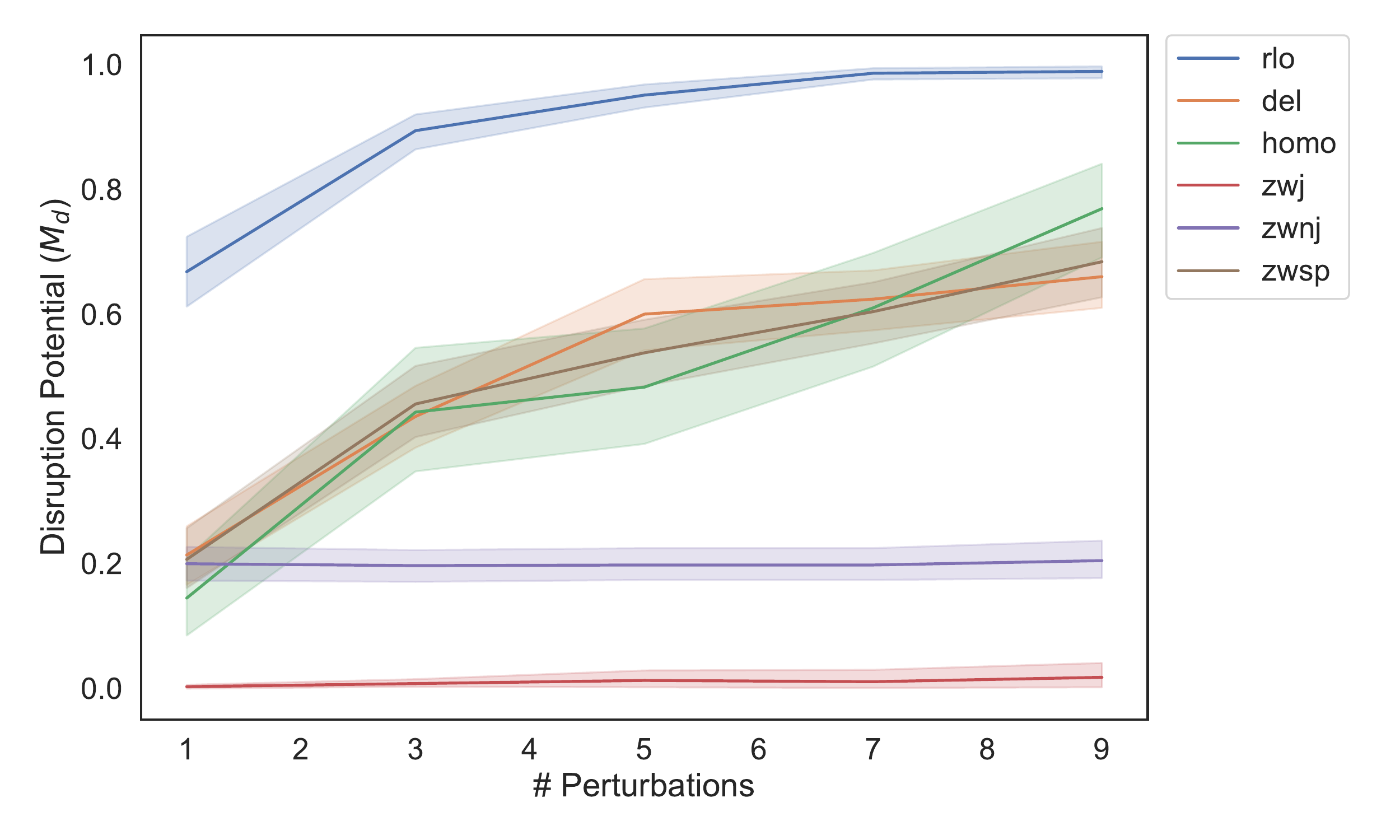}
         \caption{Google.}
     \end{subfigure}
          \begin{subfigure}[b]{0.45\linewidth}
         \centering
         \includegraphics[width=\textwidth]{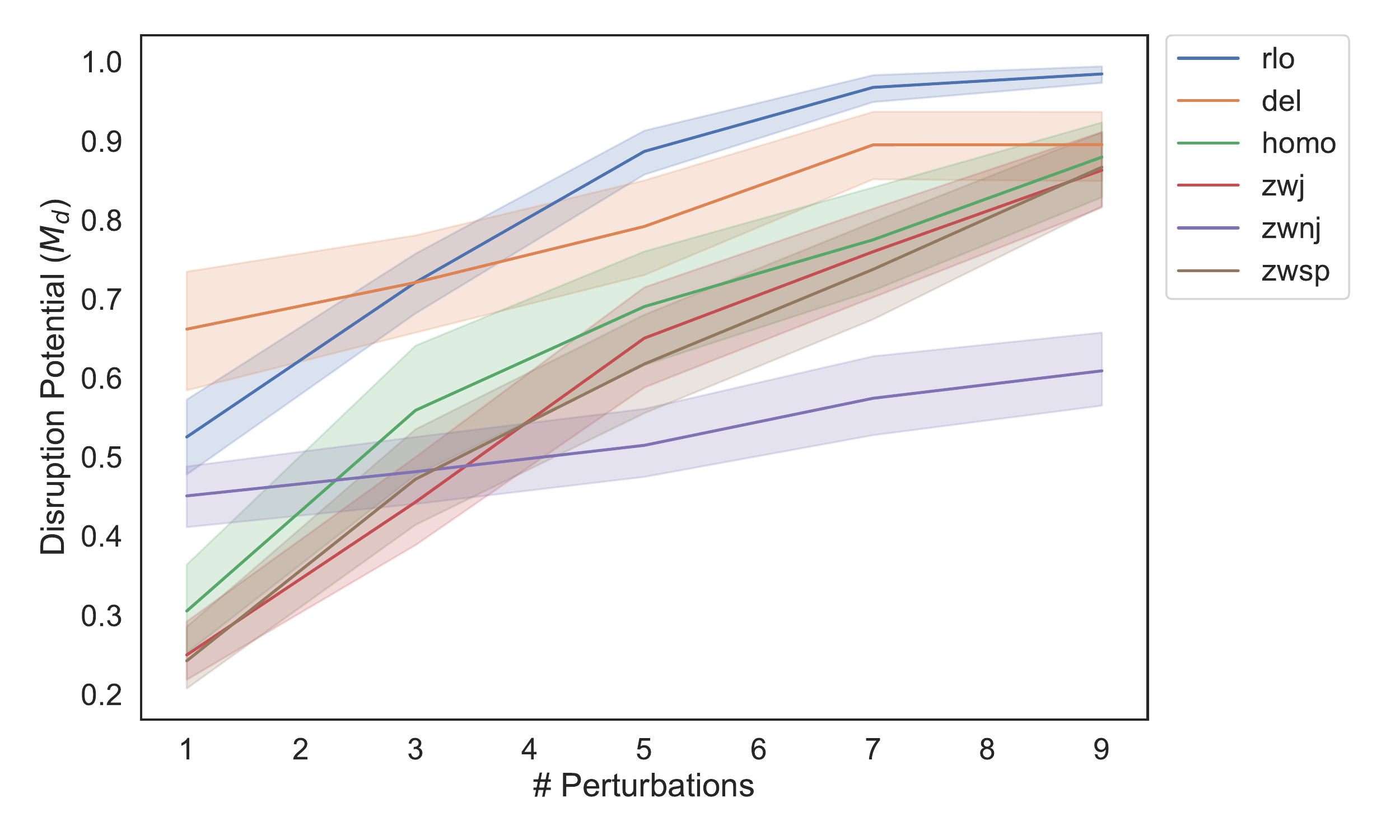}
         \caption{Bing.}
     \end{subfigure}
     
    \caption{Effect of imperceptible perturbations on different search engines at the varying amount of perturbation. The higher $M_d$, the stronger the attack.}
    \label{fig:cs1-results}
\end{figure*}

\Cref{fig:cs1-results} shows the open-internet experimental results for both Bing and Google. As expected, the results show that, in general, search engines are negatively affected by imperceptible perturbations, and as perturbations increase, performance drops significantly. Our injections' randomness can explain this phenomenon: when the perturbation is small -- \eg~one character -- its positioning might not undermine the quality of the queries. However, when inserting many characters, it is more likely that these injections destroy the semantics of victim sentences.

These results also validate that Google is resistant to ZWJ and ZWNJ injections, with a stable performance at increasing perturbation. 
Comparatively, deletion, homoglyphs, and ZWSP characters have a moderately increasing negative effect. Bidi injections display an immediate and large effect, guaranteeing a significant drop in performance with a SERP similarity close to 0.1 from the injection of only three characters. Bing, on the other hand, appears vulnerable to all the classes of evaluated attacks. Similar to Google, the Bidi attack appears the most effective.

\begin{figure}[b]
    \centering
    \includegraphics[width= 0.85\linewidth]{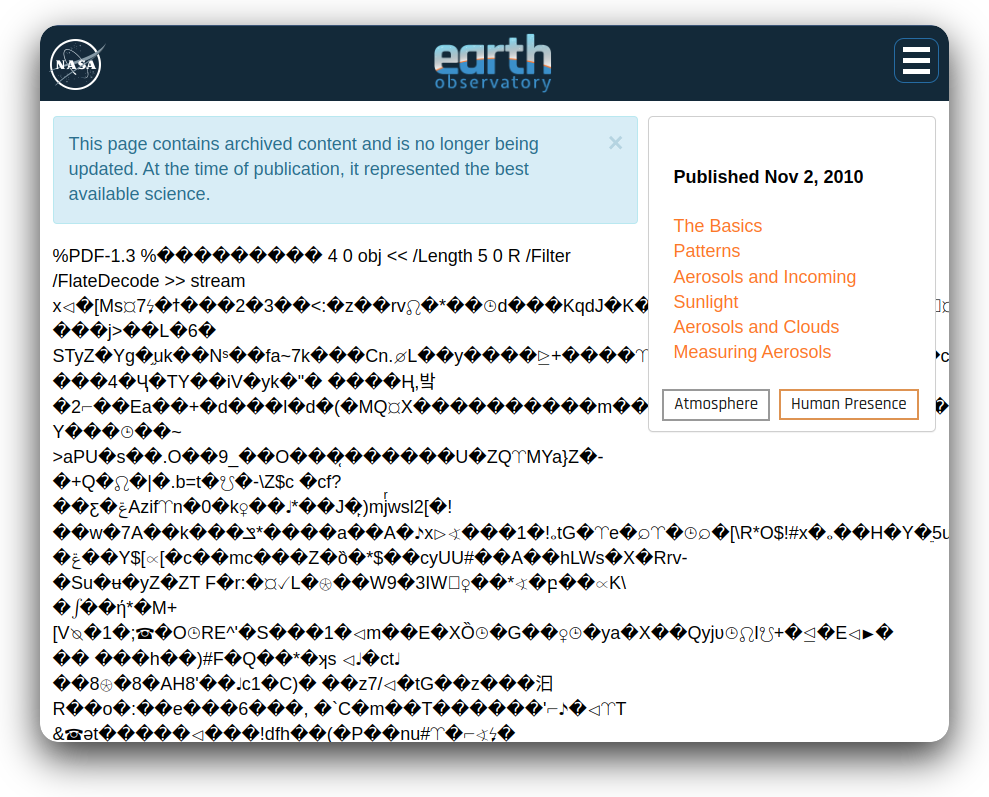}
    \caption{Example of a perturbed URL appearing in many distinct perturbed Bing queries.}
    \label{fig:nasa}
\end{figure}

We also analyzed the discrepancies between SERP from perturbed and unperturbed queries. We aimed to answer the following question: \textit{what is the nature of web pages contained in imperceptible perturbed SERPs but not in their unperturbed counterparts?} In this analysis, we combined this set of URLs returned by Google and Bing from imperceptible perturbed queries for collective analysis.

We found 15,324 and 11,080 URLs for Bing and Google, respectively, for a total of 26,404 entries. We processed such URLs with \textit{urlib}, a python library, and we extracted the following information: scheme (\eg~HTTPS), network location (\eg~google.com), URL path, URL parameters, and query. Among the 26404 URLs, we found that 507 use insecure communications (\ie~HTTP), while the other adopt HTTPS. A surprising outcome is the repetition of some network locations: in particular, we found 6,567 unique websites. \Cref{tab:common-webpages} shows the top 10 popular websites. Such results appear in the 73\% of Bing and 27\% of Google results. Similarly, the top 10 webpages are not distributed uniformly among the various attack: 36\% Bidi, 14\% deletion, 5\% homoglyph, 13\% ZWJ, 13\% ZWNJ, 19\% ZWSP.

We found some perturbed URLs repeated across distinct queries, and analyzed who the most common perturbed URL publishers were. The top 5 perturbed URLs appear in Bing responses, and all of them belong to the ``Bidi'' attack. In order, docz.net\footnote{\href{https://doczz.net/doc/7707974/e-s-q-u-e-m-a-s-d-e-s-e-n-t-i-d-o---u-n-o-q-u-e-\%E2\%80\%9C-s-e-d-u...}{doczz.net/doc/7707974/e-s-q-u-e-m-a-s-d-e-s-e-n-t-i-d-o---u-n-o-q-u-e-\textbackslash{}\%E2...}} appears 121 times in 74 distinct queries, contains a URL fragmented by hyphens, and resolves to content similarly fragmented with whitespace. Researchgate.com\footnote{\href{https://www.researchgate.net/publication/301747842_R_e_l_a_t_i_o_n_s_h_i_p_b_e_t_w_e_e_n_t_h_e_b_r_a_i_n_a_n_d_a_g_g_r_e_s_s_i_o_n}{www.researchgate.net/publication/301747842\_R\_e\_l\_a\_t\_i\_o\_n\_s\_h\_i\_p\_b\_e\_t\_w...}} appears 118 times in 79 distinct queries; this web page exhibits similar fragmentation in its URL and title. Earthobservatory.nasa.com\footnote{\href{https://earthobservatory.nasa.gov/features/Aerosols/what_are_aerosols_1999.pdf}{earthobservatory.nasa.gov/features/Aerosols/what\_are\_aerosols\_1999.pdf}} appears 80 times in 59 distinct queries, and contains a broken PDF, as show in \Cref{fig:nasa}. Next is Researchgate.com\footnote{\href{https://www.researchgate.net/publication/360221630_I_M_P_A_C_T_O_D_E_L_A_S_N_U_E_V_A_S_T_E_C_N_O_L_O_G_I_A_S_E_N_L_A_E_N_S_E_N_A_N_Z_A_-A_P_R_E_N_D_I_Z_A_J_E_DEL_INGLES_EN_LA_EDUCACION_SUPERIOR}{www.researchgate.net/publication/360221630\_I\_M\_P\_A\_C\_T\_O\_D\_E\_L\_A\_S...}} again, appearing 75 times in 58 distinct queries; this webpage exhibits a fragmented URL, title, and content. Finally, text-id.123dok.com\footnote{\href{https://text-id.123dok.com/document/zpv7334z-h-s-i-l-g-n-el-a-i-r-e-t-a-m-g-n-i-t-i-r-w-s-k-s-a-t-g-n-i-s-u-e-g-a-u-g-n-a-l-d-e-s-a-b-r-o-f-g-n-i-n-r-a-e-l-e-h-t-f-o-s-t-n-e-d-u-t-s-e-d-a-r-g-h-t-n-e-t-a-t-r-a-k-a-y-g-o-y-a-i-r-a-m-a-t-n-a-s-a-m-s-a.html}{text-id.123dok.com/document/zpv7334z-h-s-i-l-g-n-el-a-i-r-e-t-a-m-g-n-i-t-i-r...}}, appears 66 times among 51 distinct queries, and exhibits a fragmented URL and content.
\section{Search-Adjacent Machine Learning}

Beyond traditional search, there are a variety of machine learning powered systems that augment modern search engine products. In this section, we evaluate the performance of these search-adjacent machine learning tasks against imperceptible perturbations.

\subsection{Chatbot Search}

\begin{figure}[t]
    \centering
    \includesvg[width=.95\columnwidth]{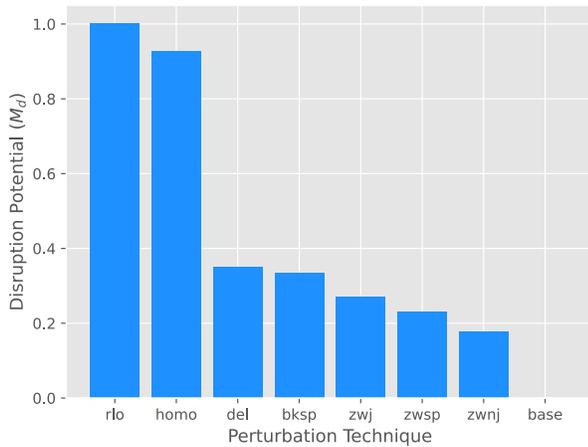}
    \caption{Comparison of Bing Chatbot web sources cited between perturbed and unperturbed inputs. The higher $M_d$, the stronger the attack.}
    \label{fig:bing-chat-disruption}
\end{figure}

\begin{figure}[b]
    \centering
    \includegraphics[width=\columnwidth]{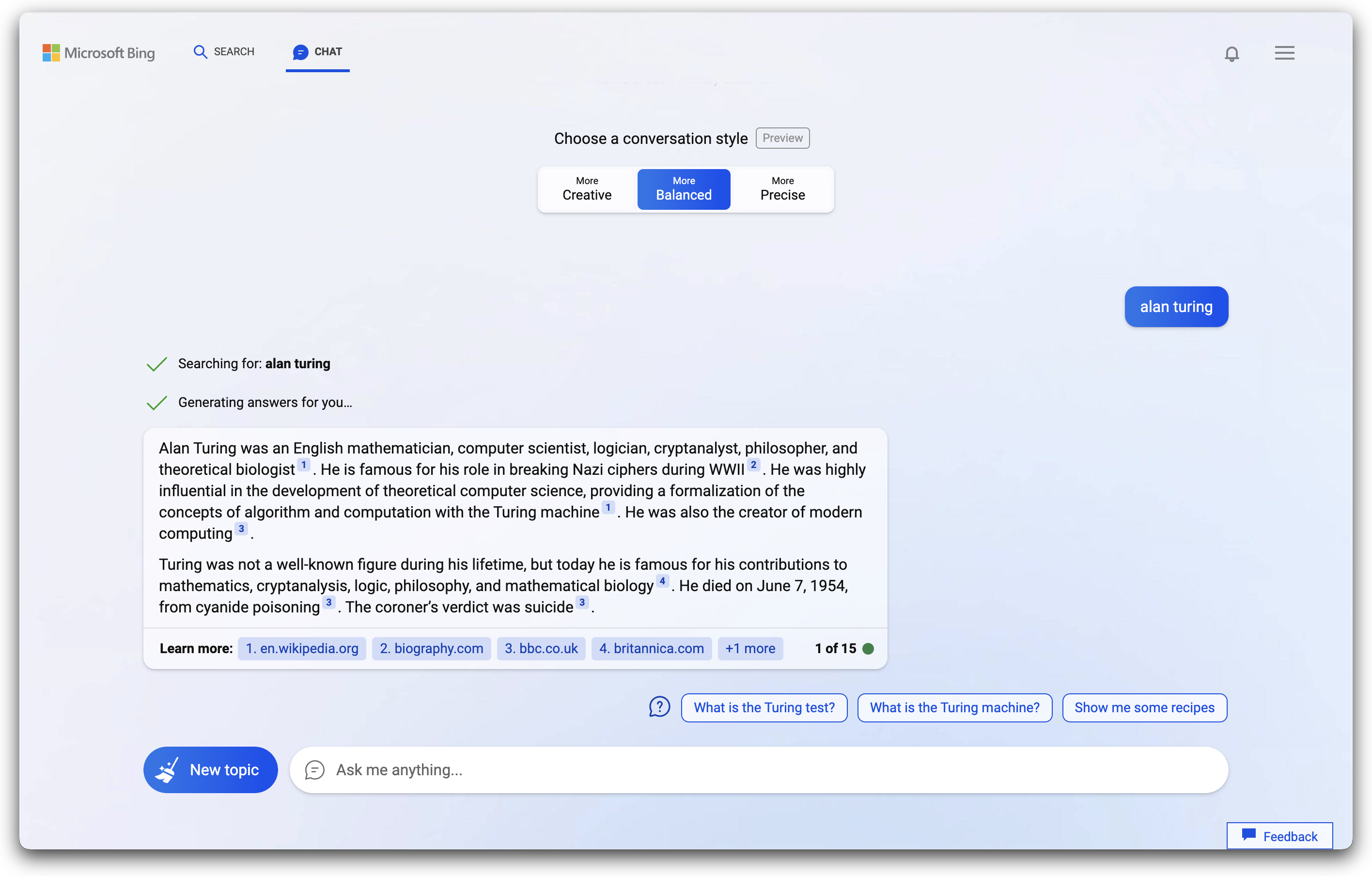}
    \caption{Bing Chatbot UI.}
    \label{fig:bing-chat-ui}
\end{figure}

\begin{figure}[t]
    \centering
    \includesvg[width=.95\columnwidth]{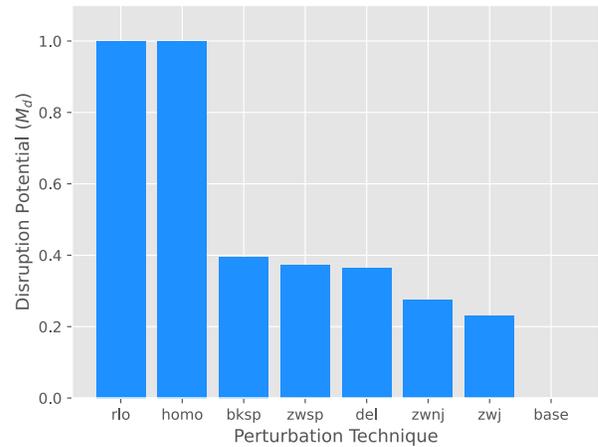}
    \caption{Comparison of Google Bard chatbot web sources cited between perturbed and unperturbed inputs. The higher $M_d$, the stronger the attack.}
    \label{fig:google-bard-disruption}
\end{figure}

Recent strides in large language models (LLMs) have led some of the largest commercial search providers to claim that the future of search will be closely coupled with LLM-driven chatbots \cite{new_bing,bard_search}. The first global-scale product to market leveraging this technology for search was Microsoft's update to Bing introducing a chatbot driven by OpenAI's GPT-4~\cite{openai2023gpt4,bing_gpt4}, followed shortly thereafter by Google's release of its competitor model Bard~\cite{bard_release}.

Given the relationship to traditional search, we were curious if both Bing's GPT-4 and Google's Bard chatbot were affected by imperceptible perturbations.

To test this, we provided the titles of the articles on our "Bad Search Wiki" as inputs to both model, repeating each inference for the different perturbation techniques listed in \Cref{tbl:perturbations}. To accomplish this, we wrote a script which directly queries each chatbot's API and captures the results. Each input was provided in the context of a fresh chat session, such that previous inputs would not affect the models' outputs.

Since the outputs of a search chatbot are different than the SERP outputs of search engines, we had to adjust the manner in which we interpret experimental results. Both the Bing and Bard conveniently provide a set of web sources with each query to serve as citations in responses generated. We compared the set of URLs returned as citations for perturbed inputs with those returned for unperturbed inputs using the disruption score $M_d$ defined in \Cref{eq:similarity}. We show this evaluation for Bing's GPT-4 model in \Cref{fig:bing-chat-disruption}. From these results, we observe that RLO and homo are the most effective perturbation techniques for disrupting chatbot response citations and are almost always successful. Each other technique, other than the control (base), also disrupted the results but had a success rate less than half that of the strongest techniques. We show the same evaluation for Google's Bard in \Cref{fig:google-bard-disruption} and observe that the results for each technique follow the same pattern as Bing but with a slightly higher attack success rate for nearly every perturbation technique.

\begin{figure}[t]
    \centering
    \includesvg[width=.95\columnwidth]{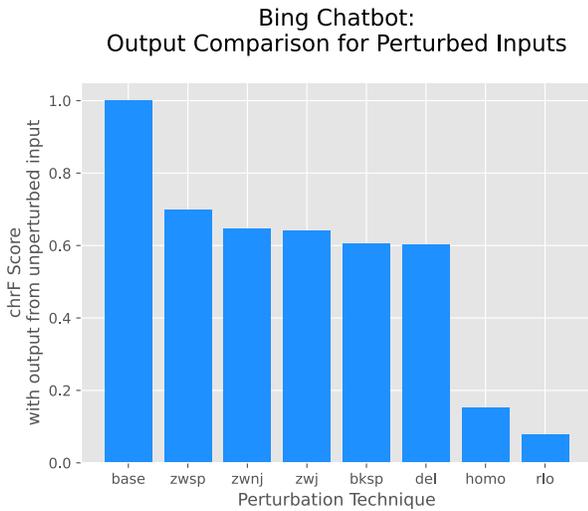}
    \caption{Comparison of Bing Chatbot text outputs between perturbed and unperturbed inputs. The lower chrF Score, the stronger the attack.}
    \label{fig:bing-chat-chrf}
\end{figure}

\begin{figure}[t]
    \centering
    \includesvg[width=.95\columnwidth]{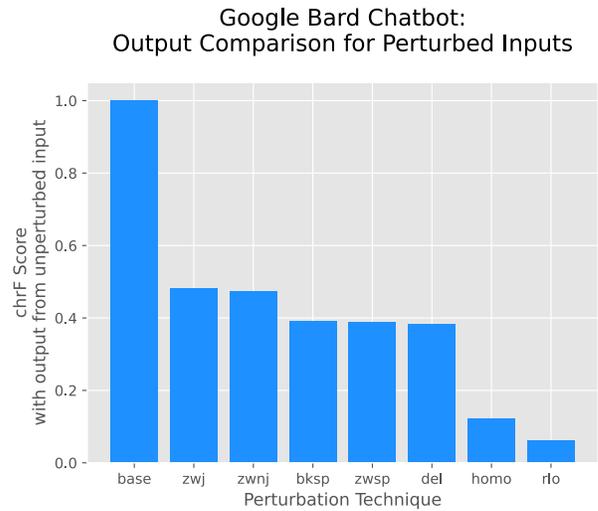}
    \caption{Comparison of Google Bard chatbot text outputs between perturbed and unperturbed inputs. The lower chrF Score, the stronger the attack.}
    \label{fig:google-bard-chrf}
\end{figure}

In addition, we wanted to evaluate the non-URL text emitted by the chatbot in the presence of imperceptible perturbations. To accomplish this, we calculated the chrF score~\cite{popovic-2015-chrf} for the perturbed model output with the unperturbed model output as the ground truth reference. In this metric, we compare only the text output and do not consider the web sources analyzed in the previous experiment. These results for Bing can be seen in \Cref{fig:bing-chat-chrf}. From these results, we see again that the Bing GPT-4 model was most affected by rlo and homo perturbation techniques, with a notable but less powerful affect on each other perturbation technique. The same evaluation for Google's Bard is show in \Cref{fig:google-bard-chrf}, and once again the trends are nearly identical to Bing's GPT-4 but with a higher attack success rate.

From these results, we can conclude that both Bing's GPT-4 chatbot and Google's Bard chatbot is highly vulnerable to manipulation via bidirectional control characters and homoglyphs, and at least somewhat vulnerable to each other perturbation technique examined.

\subsection{Summarization}\label{sub.C1-overview}

Automatic Text Summarization (ATS) is the process of reducing the length of a source of text without losing salient aspects or essential information. ATS are critical tools, especially in the Web context, where users seeking information might need to deal with many distinct sources containing enormous quantities of data~\cite{el2021automatic}. Following the outcomes shown so far, we now demonstrate that imperceptible perturbations can affect ATS. The reasons why the attack might succeed are similar to the conjecture described in the ZeW-attack~\cite{pajola2021fall}: the introduction of imperceptible perturbations affects the indexing stage of NLP pipelines; therefore, since such characters are likely not present in a regular vocabulary, obfuscated sentences will be translated into a sequence of meaningless representations.

In this work, we proposed a novel generative adversarial procedure for ATS. To the best of our knowledge, no prior works discuss adversarial attacks in ATS. Therefore, we now formalize attackers' goals and capabilities. Conceptually, attacking an ATS is similar to the more traditional \textit{evasion attack}~\cite{barreno2006can}, where adversarial examples are carefully crafted to produce a misclassification in the victims model. Similarly, in adversarial ATS, an attacker might aim to produce the following types of attack:
\begin{itemize}
    \item \textit{Targeted attack}, where the summarized text reflects only targeted properties. For instance, an attacker might aim to produce a summarization that avoids specifc concepts or, similarly, to a summarization that avoids targeted emotions (\eg~the output does not contain negative sentences);
    \item \textit{Untargeted attack}, where the unique purpose is to maximize the output degradation. This scenario can be seen as a \textit{Denial-of-Service}. 
\end{itemize}

In this work, we focus on the latter scenario, \ie~untargeted attack, and we design a novel adversarial procedure by considering the following aspects:
\begin{enumerate}
    \item \textit{Passive attack}: as perturbed samples are likely to appear in webpages, attackers do not know what summarizers will be applied, nor when the page will be summarized.
    \item \textit{Gradient-free optimization}: attackers might prefer educated guessing or heuristics to produce perturbed samples since gradient-based attack might require knowledge of victims' model~\cite{apruzzeseposition}.
    \item \textit{Imperceptible}: imperceptible perturbations maximize the stealthiness of the attack since they are by definition imperceptible~\cite{pajola2021fall, boucher2022bad}. 
\end{enumerate}

\subsubsection{The Mikado Algorithm}
We now describe the proposed attack we named ``Mikado Algorithm''. Following the observations made in Section~\ref{sub.C1-overview} and what is described by Pajola and Conti~\cite{pajola2021fall}, for the untargeted scenario, the maximum expectation of attack is found when the adversarial samples are entirely obfuscated. This means that, for Bidi, deletion, ZWJ, ZWNJ, and ZWSP, perturbed characters are inserted between each character of the given sentence; similarly, in a homoglyph attack, each sentence's character is replaced with its homoglyph counterpart. We refer to $\mathcal{O}$ as the obfuscation function that produces the strongest adversarial sample for a given input $x$ and technique $t$, with $t \in [$ bidi, deletion, homoglyph, ZWJ, ZWNJ, ZWSP$]$. Therefore, 
\begin{equation}
    x^{adv} = \mathcal{O}(x, t).
\end{equation}

The attacker might evaluate the quality of the produced adversarial sample by leveraging an external ATS $\mathcal{A}$, and an evaluation metric $\mathcal{M}$. For instance, $\mathcal{A}$ can be an online free tool or a python pre-trained library. $\mathcal{M}$ needs to measure the similarity between the summarization between $x$ and $x_adv$; in our implementation, we use the BLEU score:
\begin{equation}
\label{eq:bleu}
    m = \mathcal{M}\Bigl(\mathcal{A}(x), \mathcal{A}(x^{adv})\Bigl),
\end{equation}

with the similarity $m \in [0, 1]$, where 0 indicates that the two summarized text are completely different, and 1 are identical. Note that, for a fair comparison between $\mathcal{A}(x)$ and $\mathcal{A}(x^{adv})$, the adversarially summarized text needs to be sanitized of imperceptible perturbation. This is necessary since, otherwise, two visually identical words (one with imperceptible perturbations and the other without) will be treated as two different words from the BLEU score.

\begin{algorithm}[t]
    \begin{algorithmic}
        \caption{Mikado generation process}
        \label{alg:mikado}
        \Require $x$ \Comment{Original sample}
        \Require $t$ \Comment{Type of obfuscation}
        \Require $p$ \Comment{Size of population}
        \Require $n$ \Comment{Step of sanitization}
        \Require $b$ \Comment{Budget}
        \State $x^{adv} = \mathcal{O}(x, t)$  
        \State $m = \mathcal{M}\Bigl(\mathcal{A}(x), \mathcal{A}(x^{adv})\Bigl)$
        \While{$m \leq b$}
            \State $\textbf{x}^{cand} = \mathcal{S}(x^{adv}, p, n)$
            \State $i = \min_i \mathcal{M}\Bigl(\mathcal{A}(x), \mathcal{A}(x^{cand}_i)\Bigl)$
            \State $m = \mathcal{M}\Bigl(\mathcal{A}(x), \mathcal{A}(x^{cand}_i)\Bigl)$
            \If{$m \leq b$}
                \State $x^{adv} = x^{cand}_i$ 
            \EndIf
        \EndWhile
    \end{algorithmic}
\end{algorithm}

The produced $x^{adv}$ might contain unnecessary imperceptible perturbations that, if removed, will not alter the overall quality of the attack. We recall that while the homoglyph attack produces a sentence with the same amount of characters, this ratio is $\times 2$ when considering deletion, ZWNJ, ZWJ, and ZWSP; the Bidi algorithm, instead, produces $\times 5$ characters. To reduce the number of perturbed characters without impacting the attack's quality, we designed a genetic algorithm, given their success in NLP adversarial generations~\cite{alzantot2018generating, boucher2022bad}. The procedure is based on a sanitization function $\mathcal{S}$ that, given an adversarial sample $x^{adv}$, generates $p$ versions of the adversarial samples by randomly removing $n$ perturbed characters. We call this list of candidates $\textbf{x}^{cand}$:
\begin{equation}
    \textbf{x}^{cand} = \mathcal{S}(x^{adv}, p, n).
\end{equation}

The best candidate in the population is found as follows:
\begin{equation}
    \min_i \mathcal{M}\Bigl(\mathcal{A}(x), \mathcal{A}(x^{cand}_i)\Bigl), i\in[0, p - 1].
\end{equation}

Note that it is likely that the best candidates $x_i^{cand}$ produce an attack weaker than $x^{adv}$. Therefore, we introduce a budget parameter $b \in [0, 1]$ that indicates the maximum similarity we are willing to accept among the original and adversarial summarization:

\begin{equation}
x^{adv}=\begin{cases}
x_i^{cand}, & \text{if $\mathcal{M}\Bigl(\mathcal{A}(x), \mathcal{A}(x^{cand}_i)\Bigl) \leq b$}\\
x^{adv}, & \text{otherwise}
\end{cases}.    
\end{equation}

If $x^{adv}$ is updated, we repeat the procedure by generating a new population of sanitized candidates; on the opposite, we return $x^{adv}$.\ \ \Cref{alg:mikado} shows Mikado's pseudo-algorithm. Attackers not willing to minimize the amount of perturbation, or simply without access to any ATS, can skip the optimization procedure, and use the maximum perturbed adversarial sample. 

\subsubsection{Summarization Attack Evaluation}

\begin{table*}[t]
    \centering
    \footnotesize
    \caption{Examples of summarization failures.}
        \label{tab:tab-summ}
    \begin{tabular}{p{8cm}p{8cm}r}\toprule
         \textbf{True Summarization}& \textbf{Adversarial Summarization} & \textbf{Score}\\ \midrule
         CNN confirms that Michelle MacLaren is leaving the upcoming "Wonder Woman" movie. The movie, starring Gal Gadot in the title role of the Amazon princess, is still set for release on June 23, 2017. It's the first theatrical movie centering around the most popular female superhero.& Wanted to shoot footage of golden lassos and invisible jets. MacLaren was announced as the director of the movie in November. Warner Bros. and Michelle MacLaren have decided not to move forward with the movie. & 0.01 \\ \\
         Daniel Boykin, 33, entered the woman's home multiple times, where he took videos, photos and other data. Police found more than 90 videos and 1,500 photos of the victim on Boykin's phone and computer. The victim filed a complaint after seeing images of herself on his phone last year. & Daniel Boykin, 33, videotaped a female co-worker in a TSA-on-ly restroom. The woman filed a complaint after seeing images of herself on his phone. Boykin worked in an administrative capacity and did n't engage in public security screening. & 0.21 \\ \\
         The Americans" is one of the best series on TV. "Fresh Off the Boat" is the first sitcom with an Asian-American cast since the 1990s. "Scorpion," about a ragtag band of geniuses sent on secret missions, got a lot of hype. & The Americans is one of the best series on TV, critics say. The finale is titled March 8, 1983, when President Reagan called the Soviet Union. The first History Channel scripted series wraps up on Thursday. & 0.15\\\bottomrule
    \end{tabular}
\end{table*}

In our experiment, we aim not only to prove the ferocity of the Mikado algorithm toward a target model $\mathcal{A}$ used during the optimization process in a black-box manner but to show further that the generated adversarial samples can \textit{transfer} among different ATS. We recall that with transferability, we refer to the ability of an attack to produce adversarial samples effective in any unknown model, and not only the one used during their generation~\cite{papernot2016transferability}. In our experiment, we used the ``cnn\_dailymail'' dataset (v3.0.0)~\cite{HermannKGEKSB15, see-etal-2017-get}, consisting in a corpus of news articles written by journalists at CNN and the Daily Mail. For the attack, a random subsample of 500 instances is utilized. We considered three popular ATS trained on the ``cnn\_dailymail'' dataset, and available on \textit{Hugging Face}\footnote{\url{https://huggingface.co/}}: 
\begin{itemize}
    \item \texttt{facebook/bart-large-cnn};
    \item \texttt{sshleifer/distilbart-cnn-12-6};
    \item \texttt{sshleifer/distilbart-xsum-12-6}.
\end{itemize}

\begin{figure}[b]
    \centering
    \includegraphics[width = 0.95\linewidth]{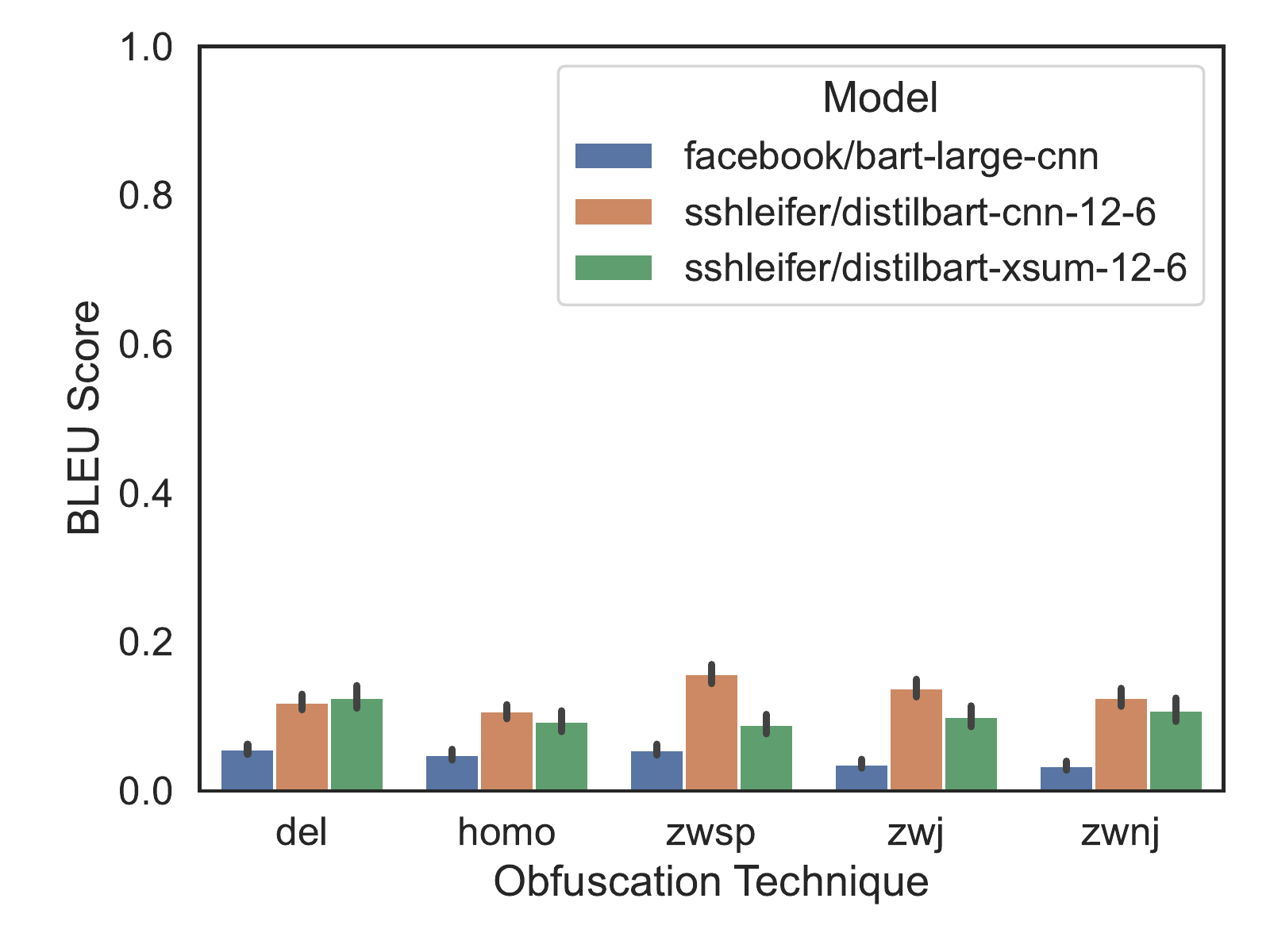}
    \caption{Attack performance varying of ATS and obfuscation technique according to \Cref{eq:bleu}. The lower the score, the stronger the attack.}
    \label{fig:summ-transferability}
\end{figure}

We use the former model in the Mikado optimization process, while the remaining models evaluate the attack transferability. 
We set $p=20$, $n=250$, and $b = 0.1$.
In our implementation, we do not test the ``Bidi attack'' since we could not guarantee a proper output sanitization; we recall that the Bidi algorithm inserts, for each transformation, multiple sequences of characters. The order of such characters might not be respected in the summarized version. 

The result shown in Figure~\ref{fig:summ-transferability} confirm our intuition: ATS are affected by imperceptible perturbations. This is true not only on the model used during the Mikado optimization but also while transferring to unknown models. 
The attack is effective, with, on average, a score equal to 0.1 for the Facebook model, corresponding to the budget $b$ limit we imposed. Similarly, the efficacy is preserved in the unknown models, with all the cases below 0.2. 

Table~\ref{tab:tab-summ} shows examples of adversarial summarized text. 
The reader might notice that when the score is close to zero, the adversarial summarization appears utterly different from the correct version. 
When the score is higher (\eg~0.2), adversarial summarizations appear to discuss the same topic as their correct versions; however, they fail to preserve the target insights.

\subsection{Plagiarism Detection}

Another family of tools that can be affected by imperceptible perturbations is plagiarism detection. Plagiarism is a fraudulent appropriation of another party's intellectual property. The goal of such a tool is, given a source of text, to spot potential third party sources that are exactly or nearly identical. Plagiarism checkers are essential in many fields to help verify the ownership of new content. Common use cases include the journalism and the scientific communities. In this setting, we consider a plagiarism checker as a classifier that returns 1 if the input content is plagiarized, and 0 if plagiarism is not detected. The attacker's goal is to evade the detection of such classifiers when using plagiarized content.

We performed experiments against Grammarly\footnote{\href{https://www.grammarly.com}{grammarly.com}}, a popular tool for grammatical corrections widely used in both academia and industry with integration into many web applications (\eg~Overleaf). Among the tools offered within Grammarly is a plagiarism checker. Since the tool does not provide APIs to interact directly with the service, we conducted the experiments manually through a web interface. For this experiment, we used 50 samples of Kaggle's Quote dataset{\footnote{\href{https://www.kaggle.com/datasets/akmittal/quotes-dataset}{kaggle.com/datasets/akmittal/quotes-dataset}}}. For each sample, we injected the maximum amount of perturbation for all the attacks we describe in \Cref{tbl:perturbations}, except deletion. The deletion attack cannot be performed since deletion control characters are actioned by the web browser before ingestion into the Grammarly interface. We measure the attack using \textit{attack success rate}, which we define in this setting as the percentage of misclassifications.

The results of the experiment
% shown in \Cref{fig:grammarly-asr} 
demonstrate the ferocity of imperceptible perturbations. In particular, all the analyzed attacks appear to be effective, producing a perfect attack success rate. In other words, Grammarly did not detect any of our imperceptibly perturbed plagiarized content as plagiarism.
\section{Discussion}

\subsection{Ethics}

Whenever novel attacks are proposed, ethics must be closely considered. We found that two commercial search engines were vulnerable to our proposed attacks, and therefore notified the maintainers of these systems prior to public release of this paper. We also paid full price for all commercial search engine APIs used to conduct the experiments described within this paper.

\subsection{Defense}

Search engine usage is typically an interactive operation. This suggests a simple mitigation that can be easily retrofitted onto existing search engines: search engines should provide some visual indication when imperceptible perturbations are detected. This could take the form of a warning message, highlighting suspected-perturbed characters in the search box, or anything else that brings visual attention to encoding level perturbations. Attack detection is straightforward: a RegEx that detect invisible characters, bidirectional control characters, deletion control characters, and selected homoglyphs would suffice.

Search engine maintainers looking for a more robust solution can build a text sanitization phase into their search pipelines. This would need to take the form of removing invisible characters, resolving control characters, and replacing homoglyphs with common characters. Homoglyph replacement is not as straightforward as the other tasks, but there are ways in which a collection of explicitly defined homoglyphs can be defined for mapping to common characters during sanitization. A precedent is given by the Punycode standard for mapping strings containing Unicode characters, such as internationalized domain names, into the subset of ASCII favored by DNS~\cite{RFC3492}.

Barring either of these defenses, each of the search engines examined, and likely future search engine implementations, will be vulnerable to this attack and provide an avenue that could support disinformation campaigns.

\begin{table*}[t]
\caption{Summary of all experimental results.\\Larger numbers / greener cells are more successful attacks.}
\label{tbl:eval-summary}
\begin{tabular}{lllllllll}
\toprule
\textbf{Target}           & \textbf{Metric} & \multicolumn{7}{l}{\textbf{Metric Change} (Relative to No Perturbation)}                                                                                                                       \\
\textbf{}                 & \textbf{}       & \textbf{zwsp}             & \textbf{zwnj}            & \textbf{zwj}             & \textbf{homo}             & \textbf{rlo}              & \textbf{bksp}             & \textbf{del}             \\ \midrule
Google Search             & $M_h$           & \cellcolor{Green!50}100\% & \cellcolor{Green!1}1\%   & \cellcolor{Green!1}2\%   & \cellcolor{Green!50}100\% & \cellcolor{Green!50}100\% & \cellcolor{Green!50}100\% & \cellcolor{Green!48}96\% \\
Google Search             & $M_s$           & \cellcolor{Green!3}6\%    & \cellcolor{Green!3}6\%   & \cellcolor{red!3}-5\%    & \cellcolor{Green!29}58\%  & \cellcolor{Green!30}59\%  & \cellcolor{red!10}-19\%   & \cellcolor{red!11}-21\%  \\
Google Search             & $M_d$ {\small($b=9$)}   & \cellcolor{Green!34}68\%  & \cellcolor{Green!11}21\% & \cellcolor{Green!1}2\%   & \cellcolor{Green!39}77\%  & \cellcolor{Green!50}99\%  & n/a                       & \cellcolor{Green!33}66\% \\
Bing Search               & $M_h$           & \cellcolor{Green!49}98\%  & \cellcolor{Green!49}98\% & \cellcolor{Green!49}98\% & \cellcolor{Green!49}98\%  & \cellcolor{Green!39}77\%  & \cellcolor{Green!29}58\%  & \cellcolor{Green!35}69\% \\
Bing Search               & $M_s$           & \cellcolor{red!6}-11\%    & \cellcolor{red!3}-5\%    & \cellcolor{red!13}-25\%  & \cellcolor{red!14}-27\%   & \cellcolor{Green!1}2\%    & \cellcolor{red!29}-58\%   & \cellcolor{red!21}-41\%  \\
Bing Search               & $M_d$ {\small($b=9$)}   & \cellcolor{Green!44}87\%  & \cellcolor{Green!31}61\% & \cellcolor{Green!43}86\% & \cellcolor{Green!44}88\%  & \cellcolor{Green!49}98\%  & n/a                       & \cellcolor{Green!45}90\% \\
Elasticsearch             & $M_h$           & \cellcolor{Green!50}99\%  & \cellcolor{Green!50}99\% & \cellcolor{Green!50}99\% & \cellcolor{Green!50}99\%  & \cellcolor{Green!50}99\%  & \cellcolor{Green!50}99\%  & \cellcolor{Green!50}99\% \\
Elasticsearch             & $M_s$           & \cellcolor{red!49}-98\%   & \cellcolor{Green!1}1\%   & \cellcolor{Green!1}1\%   & \cellcolor{Green!1}1\%    & \cellcolor{Green!1}1\%    & \cellcolor{Green!1}1\%    & \cellcolor{Green!1}1\%   \\
Bing Chatbot (GPT-4)      & $M_d$           & \cellcolor{Green!12}23\%  & \cellcolor{Green!9}18\%  & \cellcolor{Green!14}27\% & \cellcolor{Green!47}93\%  & \cellcolor{Green!50}100\% & \cellcolor{Green!17}33\%  & \cellcolor{Green!18}35\% \\
Bing Chatbot (GPT-4)      & -chrF           & \cellcolor{Green!15}30\%  & \cellcolor{Green!18}35\% & \cellcolor{Green!18}36\% & \cellcolor{Green!43}85\%  & \cellcolor{Green!46}92\%  & \cellcolor{Green!20}39\%  & \cellcolor{Green!20}40\% \\
Google Bard               & $M_d$           & \cellcolor{Green!19}37\%  & \cellcolor{Green!14}28\% & \cellcolor{Green!12}23\% & \cellcolor{Green!50}100\% & \cellcolor{Green!50}100\% & \cellcolor{Green!20}40\%  & \cellcolor{Green!18}36\% \\
Google Bard               & -chrF           & \cellcolor{Green!31}61\%  & \cellcolor{Green!27}53\% & \cellcolor{Green!26}52\% & \cellcolor{Green!44}88\%  & \cellcolor{Green!47}94\%  & \cellcolor{Green!31}61\%  & \cellcolor{Green!31}62\% \\
ATS: bart-large-cnn       & 1-BLEU    & \cellcolor{Green!47}94\%  & \cellcolor{Green!49}97\% & \cellcolor{Green!48}96\% & \cellcolor{Green!48}95\%  & n/a                       & n/a                       & \cellcolor{Green!47}94\% \\
ATS: distilbart-cnn-12-6  & 1-BLEU    & \cellcolor{Green!42}84\%  & \cellcolor{Green!44}87\% & \cellcolor{Green!43}86\% & \cellcolor{Green!45}89\%  & n/a                       & n/a                       & \cellcolor{Green!44}88\% \\
ATS: distilbart-xsum-12-6 & 1-BLEU    & \cellcolor{Green!46}91\%  & \cellcolor{Green!45}89\% & \cellcolor{Green!45}90\% & \cellcolor{Green!46}91\%  & n/a                       & n/a                       & \cellcolor{Green!44}87\% \\ \bottomrule
\end{tabular}
\vspace{1em}
\end{table*}

\section{Related Work}

\subsection{Encoding Attacks}

Text encoding manipulation has been used across many different domains to attack a myriad of systems. One such example is NLP systems, which have previously been targeted with adversarial examples formed using Unicode perturbations~\cite{boucher2022bad,pajola2021fall}. Similar techniques were used to craft the Trojan Source attack, an attack which places vulnerabilities in source code that are not rendered to human code reviewers~\cite{boucher_trojansource_2023}. Control characters have also been used to disguise the filenames and filepaths of malware~\cite{brian_krebs_right--left_2011,jakob_lell_hacking-contest_2014,microsoft_win32sirefef_2017}. In other settings, Unicode has been leveraged to create steganographic communication channels for botnets~\cite{compagno2015boten}. Lastly, homoglyphs~\cite{SimpsonMC20apwg} and invisible characters~\cite{nathaniel} have long been used to conduct phishing campaigns.

\subsection{Disinformation Campaigns}
The internet enables individuals to connect globally by powering platforms such as social networks (\eg~Facebook), e-commerce businesses (\eg~Amazon), and online forums (\eg~Reddit). 
Despite the many benefits of a globally connected internet, it also offers many opportunities to malicious actors; in particular, it allows the rapid dissemination of potentially adversarial information. Conspiracy theories, rumors, and other forms of disinformation have the potential to affect public opinion~\cite{vicario2016spreading}, which poses a particularly potent threat to democratic societies. For instance, Bessi and Ferrara~\cite{bessi2016social} observed that the presence of bots on social network platforms in the US 2016 political elections manipulated the political discussion. Later, during the COVID-19 pandemic, conspiracy campaigns were widely spread via social networks~\cite{islam2021covid}.
\section{Conclusion}

In this paper, we presented a novel attack on search engines that leverages imperceptible perturbations in text encodings to manipulate search engine results. We find that our attacks work on real-world, deployed commercial search engines as summarized in \Cref{tbl:eval-summary}. We also found that this attack successfully extends to search-adjacent machine learning models, including search chatbots, text summarization, and plagiarism detection. When exploited, this vulnerability can be used to power disinformation campaigns. Simple defenses exist in the form of visual alerts and input sanitization, and it is necessary for search engine maintainers to adopt such defenses to mitigate this risk.

\bibliography{sources}

\end{document}